%% file: main.tex
\documentclass[fleqn,usenatbib]{mnras}

\usepackage{newtxtext,newtxmath}

\usepackage[T1]{fontenc}

\DeclareRobustCommand{\VAN}[3]{#2}
\let\VANthebibliography\thebibliography
\def\thebibliography{\DeclareRobustCommand{\VAN}[3]{##3}\VANthebibliography}

\usepackage{graphicx}	
\usepackage{amsmath}	
\usepackage{multicol}        
\usepackage{bm}		
\usepackage{pdflscape}	
\usepackage{natbib}
\usepackage{color}
\usepackage{tabularx}
\usepackage{longtable}
\usepackage{bm}
\usepackage{threeparttable}
\usepackage{CJK}

\title[Galaxy-Multiplet Clustering]{Galaxy-Multiplet Clustering from DESI DR2}

\input{author}

\date{Accepted XXX. Received YYY; in original form ZZZ}

\pubyear{\the\year{}}

\begin{document}
\label{firstpage}
\pagerange{\pageref{firstpage}--\pageref{lastpage}}
\maketitle

\begin{abstract}
We present an efficient estimator for higher-order galaxy clustering using small groups of nearby galaxies, or multiplets. Using the Luminous Red Galaxy (LRG) sample from the Dark Energy Spectroscopic Instrument (DESI) Data Release 2, we identify galaxy multiplets as discrete objects and measure their cross-correlations with the general galaxy field. Our results show that the multiplets exhibit stronger clustering bias as they trace more massive dark matter halos than individual galaxies.
When comparing the observed clustering statistics with the mock catalogs generated from the N-body simulation \textsc{AbacusSummit}, we find that the mocks underpredict multiplet clustering despite reproducing the galaxy two-point auto-correlation reasonably well. This discrepancy indicates that the standard Halo Occupation Distribution (HOD) model is insufficient to describe the properties of galaxy multiplets, revealing the greater constraining power of this higher-order statistic on galaxy-halo connection and the possibility that multiplets are specific to additional assembly bias.
We demonstrate that incorporating secondary biases into the HOD model improves agreement with the observed multiplet statistics, specifically by allowing galaxies to preferentially occupy halos in denser environments. Our results highlight the potential of utilizing multiplet clustering, beyond traditional two-point correlation measurements, to break degeneracies in models describing the galaxy-dark matter connection. 
\end{abstract}




\section{Introduction} \label{sec:intro}
\input{intro}

\section{Data and Multiplet Identification} \label{sec:data}
\input{data}

\section{Correlation Methodology \& Results} \label{sec:cluster}
\input{corr}

\section{Simulations} \label{sec:abacus}
\input{simulation}

\section{Calibration for Fiber Assignment} \label{sec:calibration}
\input{calibration}

\section{The Galaxy-Dark Matter Connection} \label{sec:HOD}
\input{hod}

\section{Discussions} \label{sec:dis}
\input{dis}

\section*{Acknowledgements}
We thank the anonymous referee for the valuable comments and suggestions.
We thank Johannes Lange and Ying Zu for their insightful feedback. 
H.W. was supported by the Herchel Smith Undergraduate Science Research Program at Harvard College and the Stanford Graduate Fellowship.
D.J.E. was supported by U.S. Department of Energy grant DE-SC0007881 and as a Simons Foundation Investigator. 
This research used resources of the National Energy Research Scientific Computing Center (NERSC), a U.S. Department of Energy Office of Science User Facility located at Lawrence Berkeley National Laboratory, operated under Contract No. DE-AC02-05CH11231. 

This material is based upon work supported by the U.S. Department of Energy (DOE), Office of Science, Office of High-Energy Physics, under Contract No. DE–AC02–05CH11231, and by the National Energy Research Scientific Computing Center, a DOE Office of Science User Facility under the same contract. Additional support for DESI was provided by the U.S. National Science Foundation (NSF), Division of Astronomical Sciences under Contract No. AST-0950945 to the NSF’s National Optical-Infrared Astronomy Research Laboratory; the Science and Technology Facilities Council of the United Kingdom; the Gordon and Betty Moore Foundation; the Heising-Simons Foundation; the French Alternative Energies and Atomic Energy Commission (CEA); the National Council of Humanities, Science and Technology of Mexico (CONAHCYT); the Ministry of Science, Innovation and Universities of Spain (MICIU/AEI/10.13039/501100011033), and by the DESI Member Institutions: \url{https://www.desi.lbl.gov/collaborating-institutions}. Any opinions, findings, and conclusions or recommendations expressed in this material are those of the author(s) and do not necessarily reflect the views of the U. S. National Science Foundation, the U. S. Department of Energy, or any of the listed funding agencies.

The authors are honored to be permitted to conduct scientific research on I'oligam Du'ag (Kitt Peak), a mountain with particular significance to the Tohono O’odham Nation.

\section*{Data Availability}
Loa data will be released as part of the DESI Data Release 2 (DR2). All data points plotted in this paper are available in Zenodo at \url{https://zenodo.org/records/17648090}.



\bibliographystyle{mnras}
\bibliography{main.bib}




\bsp	
\label{lastpage}
\end{document}

%% file: author.tex
\author[Wang et al.]{Hanyue Wang,$^{1,2,3}$\thanks{E-mail: hanyuew@stanford.edu}
Daniel J. Eisenstein,$^{1}$
Jessica Nicole Aguilar,$^{4}$
Steven Ahlen,$^{5}$
Davide Bianchi,$^{6,7}$
\newauthor
David Brooks,$^{8}$
Todd Claybaugh,$^{4}$
Axel de la Macorra,$^{9}$
Arjun Dey,$^{10}$
Biprateep Dey,$^{11,12}$
Peter Doel,$^{8}$
\newauthor
Simone Ferraro,$^{4,13}$
Andreu Font-Ribera,$^{14}$
Jaime E. Forero-Romero,$^{15,16}$
Enrique Gaztañaga,$^{17,18,19}$
\newauthor
Gaston Gutierrez,$^{20}$
Klaus Honscheid,$^{21,22,23}$
Mustapha Ishak,$^{24}$
Richard Joyce,$^{10}$
Stephanie Juneau,$^{10}$
\newauthor
David Kirkby,$^{25}$
Theodore Kisner,$^{4}$
Anthony Kremin,$^{4}$
Ofer Lahav,$^{8}$
Claire Lamman,$^{23}$
\newauthor
Martin Landriau,$^{4}$
Marc Manera,$^{26,14}$
Aaron Meisner,$^{10}$
Ramon Miquel,$^{27,14}$
Eva-Maria Mueller,$^{28}$
\newauthor
Seshadri Nadathur,$^{18}$
Gustavo Niz,$^{29,30}$
Nathalie Palanque-Delabrouille,$^{31,4}$
Will J. Percival,$^{32,33,34}$
\newauthor
Francisco Prada,$^{35}$
Ignasi Pérez-Ràfols,$^{36}$
Ashley J. Ross,$^{21,37,23}$
Graziano Rossi,$^{38}$
Eusebio Sanchez,$^{39}$
\newauthor
David Schlegel,$^{4}$
Michael Schubnell,$^{40,41}$
Joseph Harry Silber,$^{4}$
David Sprayberry,$^{10}$
Gregory Tarlé,$^{41}$
\newauthor
Benjamin Alan Weaver,$^{10}$
Rongpu Zhou,$^{4}$
Hu Zou$^{42}$
\\
$^{1}$Center for Astrophysics $|$ Harvard \& Smithsonian, 60 Garden Street, Cambridge, MA 02138, USA \\
$^{2}$Department of Physics, Stanford University, 382 Via Pueblo Mall, Stanford, CA 94305, USA \\
$^{3}$Kavli Institute for Particle Astrophysics and Cosmology, Stanford University, 452 Lomita Mall, Stanford, CA 94305, USA \\
$^{4}$Lawrence Berkeley National Laboratory, 1 Cyclotron Road, Berkeley, CA 94720, USA \\
$^{5}$Department of Physics, Boston University, 590 Commonwealth Avenue, Boston, MA 02215 USA \\
$^{6}$Dipartimento di Fisica ``Aldo Pontremoli'', Universit\`a degli Studi di Milano, Via Celoria 16, I-20133 Milano, Italy \\
$^{7}$INAF-Osservatorio Astronomico di Brera, Via Brera 28, 20122 Milano, Italy \\
$^{8}$Department of Physics \& Astronomy, University College London, Gower Street, London, WC1E 6BT, UK \\
$^{9}$Instituto de F\'{\i}sica, Universidad Nacional Aut\'{o}noma de M\'{e}xico,  Circuito de la Investigaci\'{o}n Cient\'{\i}fica, Ciudad Universitaria, \\
Cd. de M\'{e}xico  C.~P.~04510,  M\'{e}xico \\
$^{10}$NSF NOIRLab, 950 N. Cherry Ave., Tucson, AZ 85719, USA \\
$^{11}$Department of Astronomy \& Astrophysics, University of Toronto, Toronto, ON M5S 3H4, Canada \\
$^{12}$Department of Physics \& Astronomy and Pittsburgh Particle Physics, Astrophysics, and Cosmology Center (PITT PACC), \\ University of Pittsburgh, 3941 O'Hara Street, Pittsburgh, PA 15260, USA \\
$^{13}$University of California, Berkeley, 110 Sproul Hall \#5800 Berkeley, CA 94720, USA \\
$^{14}$Institut de F\'{i}sica d’Altes Energies (IFAE), The Barcelona Institute of Science and Technology, Edifici Cn, Campus UAB, \\
08193, Bellaterra (Barcelona), Spain \\
$^{15}$Departamento de F\'isica, Universidad de los Andes, Cra. 1 No. 18A-10, Edificio Ip, CP 111711, Bogot\'a, Colombia \\
$^{16}$Observatorio Astron\'omico, Universidad de los Andes, Cra. 1 No. 18A-10, Edificio H, CP 111711 Bogot\'a, Colombia \\
$^{17}$Institut d'Estudis Espacials de Catalunya (IEEC), c/ Esteve Terradas 1, Edifici RDIT, Campus PMT-UPC, 08860 Castelldefels, Spain \\
$^{18}$Institute of Cosmology and Gravitation, University of Portsmouth, Dennis Sciama Building, Portsmouth, PO1 3FX, UK \\
$^{19}$Institute of Space Sciences, ICE-CSIC, Campus UAB, Carrer de Can Magrans s/n, 08913 Bellaterra, Barcelona, Spain \\
$^{20}$Fermi National Accelerator Laboratory, PO Box 500, Batavia, IL 60510, USA \\
$^{21}$Center for Cosmology and AstroParticle Physics, The Ohio State University, 191 West Woodruff Avenue, Columbus, OH 43210, USA \\
$^{22}$Department of Physics, The Ohio State University, 191 West Woodruff Avenue, Columbus, OH 43210, USA \\
$^{23}$The Ohio State University, Columbus, 43210 OH, USA \\
$^{24}$Department of Physics, The University of Texas at Dallas, 800 W. Campbell Rd., Richardson, TX 75080, USA \\
$^{25}$Department of Physics and Astronomy, University of California, Irvine, 92697, USA \\
$^{26}$Departament de F\'{i}sica, Serra H\'{u}nter, Universitat Aut\`{o}noma de Barcelona, 08193 Bellaterra (Barcelona), Spain \\
$^{27}$Instituci\'{o} Catalana de Recerca i Estudis Avan\c{c}ats, Passeig de Llu\'{\i}s Companys, 23, 08010 Barcelona, Spain \\
$^{28}$Department of Physics and Astronomy, University of Sussex, Brighton BN1 9QH, U.K \\
$^{29}$Departamento de F\'{\i}sica, DCI-Campus Le\'{o}n, Universidad de Guanajuato, Loma del Bosque 103, Le\'{o}n, Guanajuato C.~P.~37150, M\'{e}xico \\
$^{30}$Instituto Avanzado de Cosmolog\'{\i}a A.~C., San Marcos 11 - Atenas 202. Magdalena Contreras. Ciudad de M\'{e}xico C.~P.~10720, M\'{e}xico \\
$^{31}$IRFU, CEA, Universit\'{e} Paris-Saclay, F-91191 Gif-sur-Yvette, France \\
$^{32}$Department of Physics and Astronomy, University of Waterloo, 200 University Ave W, Waterloo, ON N2L 3G1, Canada \\
$^{33}$Perimeter Institute for Theoretical Physics, 31 Caroline St. North, Waterloo, ON N2L 2Y5, Canada \\
$^{34}$Waterloo Centre for Astrophysics, University of Waterloo, 200 University Ave W, Waterloo, ON N2L 3G1, Canada \\
$^{35}$Instituto de Astrof\'{i}sica de Andaluc\'{i}a (CSIC), Glorieta de la Astronom\'{i}a, s/n, E-18008 Granada, Spain \\
$^{36}$Departament de F\'isica, EEBE, Universitat Polit\`ecnica de Catalunya, c/Eduard Maristany 10, 08930 Barcelona, Spain \\
$^{37}$Department of Astronomy, The Ohio State University, 4055 McPherson Laboratory, 140 W 18th Avenue, Columbus, OH 43210, USA \\
$^{38}$Department of Physics and Astronomy, Sejong University, 209 Neungdong-ro, Gwangjin-gu, Seoul 05006, Republic of Korea \\
$^{39}$CIEMAT, Avenida Complutense 40, E-28040 Madrid, Spain \\
$^{40}$Department of Physics, University of Michigan, 450 Church Street, Ann Arbor, MI 48109, USA \\
$^{41}$University of Michigan, 500 S. State Street, Ann Arbor, MI 48109, USA \\
$^{42}$National Astronomical Observatories, Chinese Academy of Sciences, A20 Datun Road, Chaoyang District, Beijing, 100101, P.~R.~China
}

%% file: intro.tex
Galaxy clustering encodes valuable information about large-scale structure and cosmology. 
Since galaxies are biased tracers of the underlying dark-matter density field, measurements of their clustering properties provide critical constraints on both cosmological parameters and the physics of galaxy formation, specifically the relationship between galaxies and dark matter halos (see \cite{Wechsler&Tinker} for a review). 
With the advent of advanced sky surveys, we are entering an era of high-precision studies of galaxy clustering, where galaxy distribution needs to be effectively quantified to capture the full complexity of galaxy bias and deepen our understanding of cosmology and galaxy–dark matter connection.

The most commonly-used statistical tool to quantify galaxy clustering is the two-point correlation function (2PCF, \cite{PCF_book, 2PCF}), which should be sufficient to describe the spatial matter distribution of the universe if the matter density fluctuations constitute a Gaussian random field \citep{Gaussian-random-field}. However, nonlinear gravitational instabilities and the complex physics of galaxy formation produce non-Gaussian signatures in the galaxy distribution measured today \citep{3PCF_basics}. Therefore, higher-order statistics are necessary to complement the information provided by the 2PCF when studying late-time galaxy distribution. 
Measurements of higher-order statistics using galaxy catalogs have a long-standing history. 
For example, there has been a wealth of studies using the three-point functions (3PCF) to constrain the galaxy-halo connection and to break degeneracies between cosmology and galaxy bias with both simulations and observations \citep{Gaztanaga_3PCF, Kulkarni_3PCF, Martin_3PCF, SDSS_3PCF, 3PCF-HOD2, Yuan_2018, Zhang_3PCF}. 
However, it is computationally expensive to calculate higher-order correlation functions, especially with a growing sample size from current and upcoming galaxy surveys. 

Therefore, in this work, we use galaxy multiplets, small groups of nearby galaxies, to study higher-order clustering by cross-correlating them with the full galaxy sample. Unlike galaxy groups or clusters, multiplets are not necessarily virialized, but their physical configurations occupy the regime of gravitational nonlinearities and thus contain rich information. We expect multiplets to trace dark matter differently than individual galaxies, which can be probed through their clustering patterns. 
In practice, we identify galaxy multiplets based on their spatial information, treat multiplets of $n$ galaxies as discrete objects, and measure the cross-correlation between the multiplet field and the general galaxy field. This cross-correlation is related to an $n+1$ point correlation function in the squeezed limit, referring to the configuration in which one of the vertices is significantly more separated from the rest. Through this cross-correlation between the multiplet field and the general galaxy field, the computationally expensive multipoint calculation is reduced to two separate pair-finding problems, while preserving the information from higher-order correlations and small-scale non-Gaussianity.


This approach extends the idea of the squeezed 3PCF framework introduced in \cite{3PCF2}, which focused on close galaxy pairs, with several modifications. As \cite{3PCF2} uses all available galaxy pairs to compute the cross-correlation, the measurement is effectively equivalent to a standard squeezed 3PCF. In this work, however, we explicitly separate the multiplets into systems of distinct size (i.e., pairs, triplets, and quads), so that subgroups of higher-multiplicity systems do not enter lower-order analyses. For example, a triplet can be viewed as three different pairs, but they are not included in the galaxy-pair cross-correlation calculations. Consequently, our measurement differs from the standard framework of squeezed higher-order correlations, but it remains closely related and retains higher-order information in the multiplet configurations. 
We choose this separation because systems of different multiplicities are physically distinct tracers of dark matter, and measuring the cross-correlation functions separately provides access to their physical properties directly.

Multiplets are physically interesting systems that can probe various aspects of the large-scale structure. While multiplet orientations have been used in the perspective of intrinsic alignment to study the large-scale tidal field \citep{Lamman}, here we focus on their clustering properties. The clustering of multiplets is directly connected to the underlying matter density field, providing key information about nonlinear structure formation and the galaxy-halo connection. 
While reconstructing galaxy formation through the combination of large-scale matter distribution and galaxy bias remains uncertain, the multipoint correlation functions measured with multiplets will likely probe how galaxies trace dark matter in novel ways.

We explore this new estimator of multiplet clustering using the Luminous Red Galaxy (LRG) sample from the Dark Energy Spectroscopic Instrument \citep[DESI,][]{DESI1, DESI2, DESI2022, Silber, Miler, Poppett}. We develop an efficient algorithm to identify galaxy multiplets (specifically pairs, triplets, and quads) and measure the galaxy-multiplet cross-correlation functions by cross-correlating the multiplet field with the general galaxy field.
To evaluate the effectiveness of this higher-order statistic, we compare the measured multiplet clustering with predictions from mock catalogs, demonstrating its potential to probe assembly bias and to place tighter constraints on the complex galaxy-dark matter connection.


This paper is structured as follows. In Section~\ref{sec:data}, we describe the data sample used for the study and outline the identification of galaxy multiplets. In Section~\ref{sec:cluster}, we introduce the correlation methodology for measuring multiplet clustering and present the corresponding results. We introduce the DESI mock catalogs in Section~\ref{sec:abacus} and discuss the calibration for survey systematics in Section~\ref{sec:calibration}. In Section~\ref{sec:HOD}, we compare the clustering measurements from the mocks to observations and explore the implications for the galaxy–halo connection. 
Finally, we summarize and discuss the results further in Section~\ref{sec:dis}. Throughout the paper, we assume $H_{0}=70\,\rm km/s/Mpc$.

%% file: data.tex
\subsection{The DESI LRG Sample}

DESI started the main survey observation in May 2021 after a period of survey validation \citep{SV_KP}. 
This study makes use of the DESI Year 3 dataset, collected through April 2024, with processing via the ``Loa'' version of the data compilation using the DESI spectroscopic reduction pipeline \citep{spec_pipeline} and the redshift estimation pipeline \citep[\textsc{Redrock},][]{z_estimate, Redrock.Bailey.2024}. It will be released as part of the DESI Data Release 2 (DR2) in a manner similar to the Early Data Release and the Data Release 1 \citep[DR1,][]{DR1}. 

LRGs are an important class of galaxies for large-scale structure studies and are specifically selected in this work for two reasons. First, the sample is relatively complete due to their relatively high priority in the DESI fiber assignment procedure; and second, they are highly biased tracers of the large-scale structure, offering opportunities to probe high-density environments.

LRG target selection \citep{LRG} is based on Legacy Imaging Surveys Data Release 9 \citep{LS, BASS}, specifically the $g, r, z$ optical bands, and the Wide-field Infrared Survey Explorer \citep[\textit{WISE},][]{WISE} photometry. Additional information on DESI's observation planning and target selection is available in \citet{Schlafly2023} and \cite{Myers2023}.
For the clustering analysis, we use LRGs from the DR2 large-scale structure (LSS) catalog \citep{LSS}, which incorporates details of target selection. Compared to DR1, the fiber-assignment completeness has increased by $19\%$ \citep{Y3_results}, which is important for our study, as missing galaxies in the identified galaxy multiplets due to fiber-assignment incompleteness can bias the calculation of multiplet cross-correlations. 

\begin{figure}
    \centering
    \includegraphics[width=0.9\columnwidth]{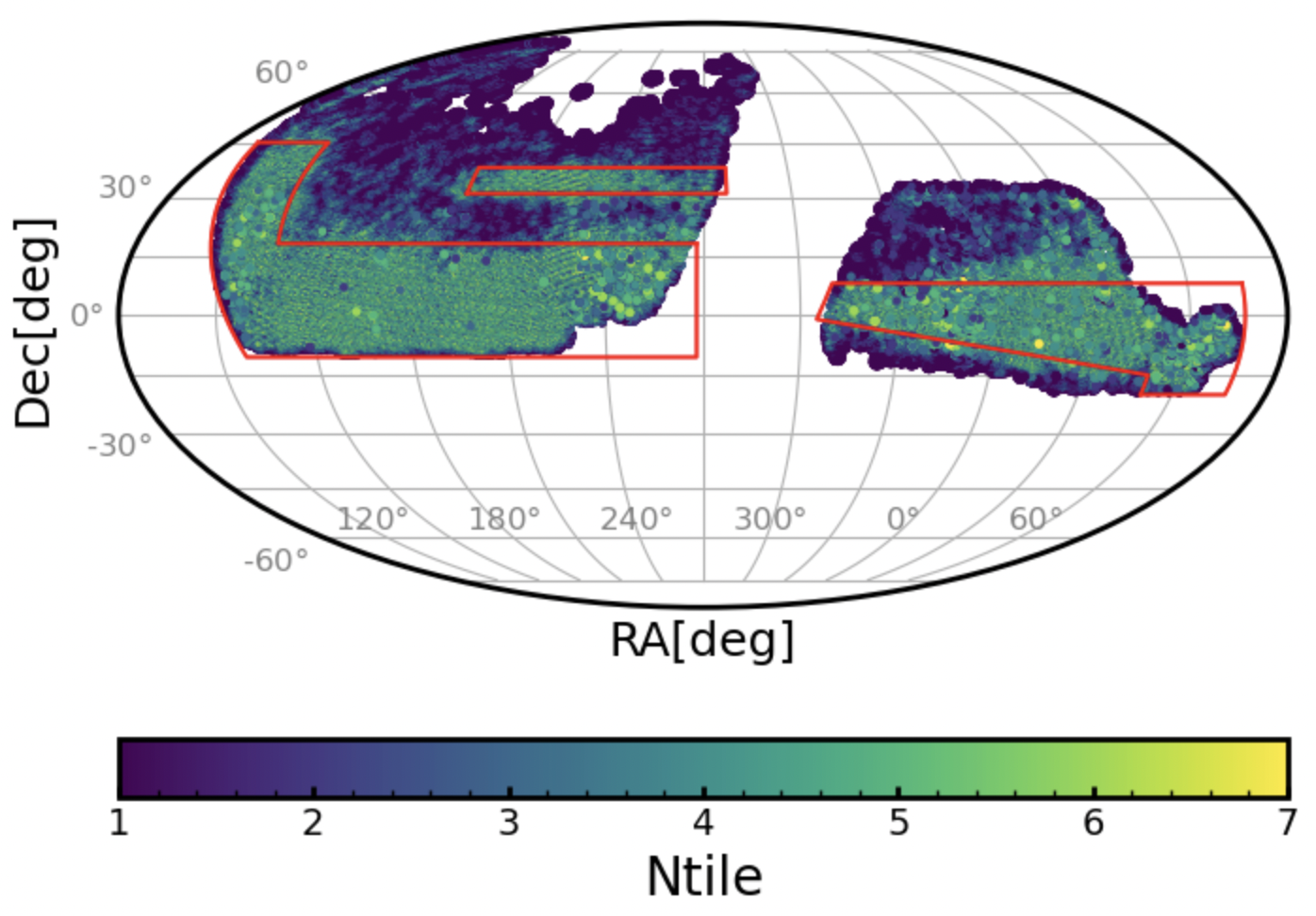}
    \caption{The footprint of the LRGs from DESI DR2, color-coded by \texttt{NTILE}. The red contour marks the region with relatively high completeness that our study focuses on. }
    \label{fig:footprint}
\end{figure}

Fig.~\ref{fig:footprint} shows the footprint of the LRGs color-coded by \texttt{NTILE}, which describes the number of times a certain portion of the sky has been observed. We choose to focus on a region with a relatively large value of \texttt{NTILE} (and then high completeness) to avoid missing galaxies in the identified multiplets and thus to reduce measurement bias in clustering statistics. This selected region, which covers approximately $5700\,\rm deg^2$ with a fiber assignment completeness of approximately $90\%$, is marked with the red contour in Fig.~\ref{fig:footprint}.

\subsection{Multiplet-Identification Algorithm}
\label{sec: cluster-finding}

The galaxy multiplets are identified using their spatial information. 
For each galaxy, we calculate its position on the cartesian coordinate based on its angular coordinates RA and Dec, and its spectroscopic redshift $z$. We utilize a flat cosmological model with matter density $\Omega_m=0.3$ and cosmological constant density $\Omega_\Lambda=0.7$ to compute the comoving distance using the redshift. With the cartesian coordinates, we look for close galaxy pairs with a separation less than $10\,\rm Mpc$ and then further select the pairs based on a transverse separation threshold of $1.5\,\rm Mpc$. The line-of-sight (LOS) separation is chosen to be much larger than the transverse separation to account for redshift-space distortions, especially the ``Fingers of God'' effect. We calibrate the transverse separation as the length of the transverse separation vector, which is calculated by subtracting the dot product of the separation vector between the galaxy pair and the vector pointing to their midpoint from the separation vector. After obtaining the pairs, we utilize a friends-of-friends algorithm to link the pairs together by assigning the same group number to all pairs containing the same galaxy, forming the catalog of galaxy multiplets. The remaining galaxies that are not part of any larger multiplet are classified as singlets.

\begin{figure}
    \centering
    \includegraphics[width=0.9\columnwidth]{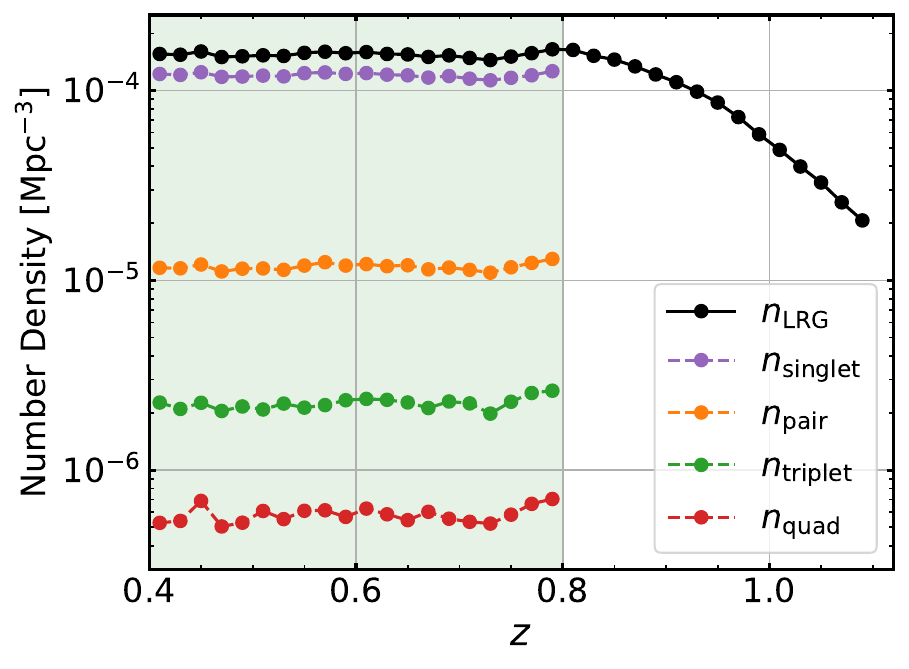}
    \caption{The redshift distributions of the LRGs and the multiplets used as tracers in the clustering analysis. The shaded area marks the selected redshift range for this work, where the number densities of all tracers are approximately constant.}
    \label{fig:n}
\end{figure}

Fig.~\ref{fig:n} shows the redshift distributions of the LRGs and the multiplets (singlets, pairs, triplets, and quads) used as tracers in this work.
At $z > 0.8$, the LRG density drops rapidly due to the flux limit of the sample in the imaging. 
To focus on the regime where the number density remains approximately constant, we restrict the analysis to LRGs in the redshift range of $0.4 < z < 0.8$, which also corresponds to the first two LRG bins in the DESI 2024 cosmological analysis \citep{DESI2024_ii, DESI2024_v, DESI2024_vii}. The redshift distribution of the multiplets demonstrates that there is little redshift evolution within this range for all tracers used in the clustering analysis. 

To summarize some statistics, from the total 1603265 LRGs in a sky area of roughly $5700 \,\rm deg^{2}$, we find 230863 galaxy pairs, which combine into 155267 multiplets of minimum size two using the friends-of-friends method. Among all the LRGs, about $22.3\%$ of galaxies reside in larger multiplets while the others remain as singlets; among all the galaxies grouped as multiplets, about $68.6\%$ of them appear in pairs. In total, we identify 122588 pairs, 23508 triplets, and 6092 quads. 
It is worth noting that as we approach larger galaxy multiplets, their number densities decrease significantly.

%% file: corr.tex
\subsection{Clustering Statistics}

In this section, we present the analytical tools used in our work, including the pair-count-based estimators for the auto-correlation and the galaxy-multiplet cross-correlation functions, together with the method for error estimation.

The calculation of correlation functions requires ``random'' catalogs, which contain samples of random sky positions that directly represent the survey data in sky geometry and redshift distribution. 
High-quality random catalogs are essential for estimators to properly represent the correlation function, as the measurements depend sensitively on the randoms to accurately reflect spatial and redshift-dependent selection effects in the actual survey data. DESI randoms are processed through the same pipeline as the LSS catalogs and are standardized to a number density of $2500\,\rm deg^{-2}$.

To ensure that clustering measurements are unbiased from survey specifics, each object in the DESI LSS and random catalogs is assigned a weight, as detailed in \cite{DESI2024_ii}. 
Briefly, each observed target in the data catalog is assigned a total weight to account for the DESI selection function. This total weight is a product of three components: the completeness weight \texttt{WEIGHT\_COMP}, the redshift failure weight \texttt{WEIGHT\_ZFAIL}, and the systematic weight \texttt{WEIGHT\_SYS}. Specifically, \texttt{WEIGHT\_COMP} describes how well-sampled the targets are in a given region of the sky. It is primarily correlated with the $\texttt{NTILE}$ parameter and the geometry of DESI's fiber positioners on the focal plane, accounting for fiber assignment incompleteness and fiber collisions \citep{DESI2023b}. 
Then, \texttt{WEIGHT\_SYS} corrects for variations in the imaging data quality, and \texttt{WEIGHT\_ZFAIL} accounts for the variations in accuracy and precision of redshift estimation in particular DESI observations. 
Weights for objects in the random catalogs are assigned from randomly chosen galaxies in the data catalogs, along with their associated redshifts, to obtain the same redshift distribution as the data. 
Ultimately, the estimator for the correlation function adopts the normalized weighted pair counts, with each pair weighted by the product of the individual weights assigned to the two objects involved.

With the DESI LSS and random catalogs, we start with the galaxy-galaxy auto-correlation by computing the anisotropic two-point correlation function $\xi (r_{p}, \pi)$ with the widely used Landy-Szalay estimator \citep{Landy}: 
\begin{equation}
    \xi\left(r_{p}, \pi\right)=\frac{DD\left(r_{p}, \pi\right)-2DR\left(r_{p}, \pi\right)+RR\left(r_{p}, \pi\right)}{RR\left(r_{p}, \pi\right)},
\end{equation}
where $r_p$ is the transverse distance and $\pi$ is the distance along the LOS. $DD\left(r_{p},\pi\right)$, $RR\left(r_{p},\pi\right)$, and $DR\left(r_{p},\pi\right)$ are the normalized weighted pair counts between galaxy-galaxy, random-random, and galaxy-random catalogs in each particular $(r_{p},\pi)$ bin. Specifically, we use $14$ $r_{p}$ bins spanning the interval $\left[0.01, 30\right]\,\rm Mpc$ with a maximum LOS separation of $\pi_{\rm max}=60\,\rm Mpc$.

To compute the galaxy-multiplet cross-correlations, we treat multiplets of different sizes as discrete objects at the average angular coordinates and redshift of their member galaxies, and cross-correlate them with the full galaxy sample. Because it is non-trivial to construct random catalogs for the multiplets, we adopt the asymmetric Davis-Peebles estimator:
\begin{equation}
    \xi\left(r_p, \pi\right)=\frac{D_{m}D_{g}\left(r_p, \pi\right)}{D_{m}R_{g}\left(r_p, \pi\right)}-1,
\end{equation}
where $D_{m}$ denotes the data catalog of multiplets (i.e., pairs, triples, and quads), and $D_{g}$ and $R_{g}$ correspond to the data and random catalogs for the entire selected DESI LRG sample. Each object in $D_{g}$ and $R_{g}$ is weighted in the same way as in the auto-correlation calculation, following the standard DESI procedure described above. All multiplets in $D_{m}$ are assigned a weight of $1$, since their multiplicity introduces more complexity in the weighting due to biases from incompleteness and other systematics.

The projected correlation function is then obtained by integrating the anisotropic 2D correlation function $\xi(r_{p}, \pi)$ along the LOS,
\begin{equation}
    w_{p}\left(r_p\right)=\int_{-\pi_{\rm max}}^{\pi_{\rm max}}\xi\left(r_p,\pi\right)\ d\pi.
\end{equation}
As the projected correlation marginalizes over the direction along the LOS, it is less informative than the full-shape $\xi(r_p, \pi)$. Still, we choose to use $w_{p}$ in our main analysis because it is easy to visualize as a 1D function and avoids the complexity of modeling galaxy velocities.

All pair-counting calculations in this work are performed with the python package \texttt{Corrfunc} \citep{Corrfunc}. 

Finally, to estimate the uncertainties in the projected correlations, we divide the catalogs of galaxy multiplets into $50$ spatial subsamples but keep the $D_{g}$ and $R_{g}$ catalogs in their original form, repeat the calculations for $w_p$ while dropping one multiplet subsample at a time, and then compute the error bars with jackknife error estimation:
\begin{equation}
    \text{Var}\left(x\right)=\frac{n-1}{n}\sum_{i=1}^{n}{(\bar{x}_i-\bar{x}_{\rm jack})}^2,~ \bar{x}_{\rm jack}=\frac{1}{n}\sum_{i=1}^{n}\bar{x}_{i},
\end{equation}
where $\bar{x}_{i}$ is the $i$-th jackknife replicate calculated by dropping the $i$-th multiplet subsample, and $n=50$ is the total number of jackknife regions. 

\subsection{Results}

\begin{figure}
    \centering
    \includegraphics[width=0.9\columnwidth]{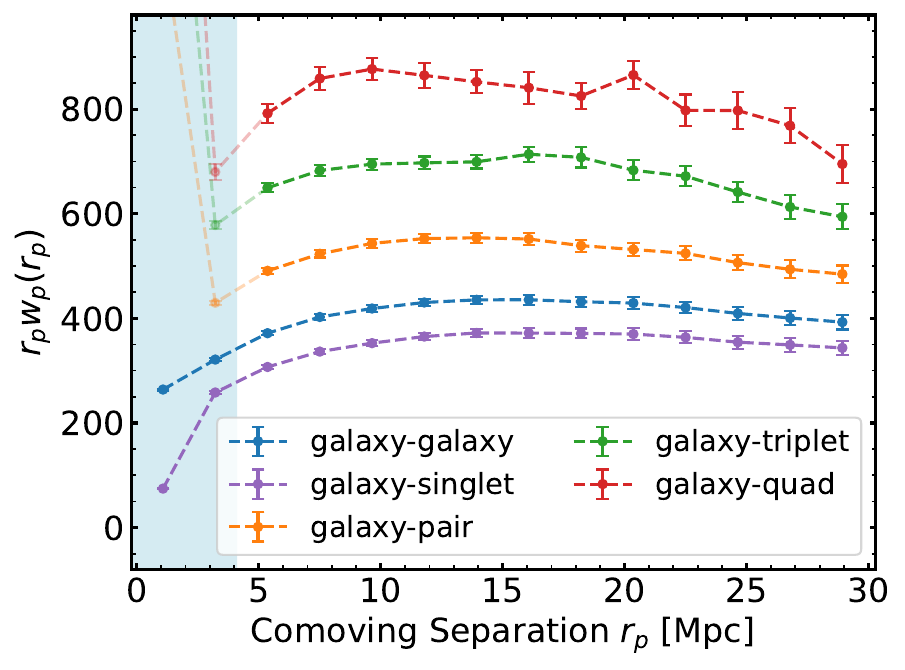}
    \caption{Comparison between the projected functions for the galaxy two-point auto-correlation and the cross-correlations with singlets, pairs, triplets, and quads.}
    \label{fig:cross-corr}
\end{figure}

Following the methodology, we present the clustering measurements from the LRGs. Fig.~\ref{fig:cross-corr} shows the galaxy-galaxy auto-correlation, along with the cross-correlations between galaxy singlets/multiplets (specifically pairs, triplets, and quads) and the general galaxy field, with the error bars estimated using the jackknife method. 
Singlets here refer to galaxies that do not belong to any larger multiplet, so their cross-correlation with the full galaxy sample is expected to be lower than the two-point auto-correlation. Meanwhile, the multiplets exhibit stronger clustering bias as they trace more massive dark matter halos. 
To ensure reliable statistics and spatial coverage, we require the multiplets to be sufficiently abundant across the analysis field. For instance, the cross-correlation with quads shows more fluctuations on large scales ($10\,\text{Mpc} < r_{p} < 30\,\text{Mpc}$) compared to those with smaller multiplets, reflecting their sensitivity to cosmic variance. Therefore, we limit the multiplet analysis to systems with up to four galaxies.

On small scales, specifically $r_{p}\lesssim 2\,\rm Mpc$, we note that the cross-correlations with galaxy multiplets increase sharply. This is because our multiplet identification algorithm sets the threshold for galaxy pairs to be $1.5\,\rm Mpc$ along the projected direction, which leads to these over-clustering signatures observed at the scale of individual galaxy multiplet. Meanwhile, the galaxy-singlet cross-correlation dips further below the auto-correlation because most small-scale clustering signals arise from the correlation between satellite galaxies and their hosts (the one-halo term in the literature), which have been largely excluded from the singlet selection. As a result, we exclude measurements for the multiplets on small scales ($r_p< 4\,\rm Mpc$, shown as the shaded area in Fig.~\ref{fig:cross-corr}), but only retain small-scale information in the auto-correlation. 

In general, multiplets trace the underlying dark matter field differently from individual galaxies. We observe excess clustering for higher-order multiplets, consistent with their association with more massive dark matter halos.
Relative to the auto-correlation, the cross-correlation with singlets has approximately $0.86$ of the amplitude, while the cross-correlations with pairs, triplets, and quads are boosted by factors of $\sim 1.25$, $1.60$, and $1.94$, respectively, demonstrating that larger multiplets are more biased tracers of the dark matter density field. 
These statistics will be valuable for comparison with the clustering properties constructed from simulations in the following sections.

%% file: simulation.tex
To connect observations with theoretical predictions, it is necessary to model the underlying dark matter density field of the universe and construct mock galaxy catalogs on top of it following some galaxy-dark matter connection model. In this section, we describe the construction of the official DESI mocks and outline the key components used to generate the mock galaxy catalogs for our analysis.

\subsection{AbacusSummit}
The DESI mocks are built with the dark matter density fields from the \textsc{AbacusSummit} simulation suite, a set of large cosmological simulations using the high-performance, high accuracy \textsc{Abacus} N-body code \citep{AbacusSummit, Abacus}. 
These simulations were specifically developed to satisfy the Cosmological Simulation Requirements of DESI.
\textsc{AbacusSummit} includes $25$ base simulations at Planck 2018 $\Lambda\rm CDM$ cosmology \citep{Planck2018}, each containing $6912^{3}\approx330$ billion particles with a particle mass of $2\times10^{9}\,h^{-1}M_{\odot}$ in a periodic box of volume $(2\,h^{-1}\rm Gpc)^{3}$ \footnote{For more details, see \url{https://abacussummit.readthedocs.io/en/latest/abacussummit.html}.}.

\subsection{Halo Occupation Distribution (HOD)}
\label{HOD_model}
To connect the galaxy distribution to the dark matter density field and to generate mock galaxy catalogs using halo catalogs from \textsc{AbacusSummit}, we employ the important framework of Halo Occupation Distribution (HOD). HOD models describe how galaxies populate dark matter halos as a function of different halo properties \citep{HOD_first}. Statistically, the HOD formalism models the conditional probability $P(n|\mathbf{X}_{h})$ that a halo with properties $\mathbf{X}_{h}$ hosts $n$ galaxies of a given type. In addition to predicting galaxy counts, HOD models also characterize the spatial and velocity distributions of galaxies relative to dark matter within halos. 

In its simplest implementation, the vanilla HOD model distinguishes between central and satellite galaxies based on their different formation mechanisms and assumes that halo mass is the only relevant parameter among all halo properties when assigning galaxies to halos \citep{Zheng_2005}. Meanwhile, extensive research using both observations and simulations has investigated secondary influences on galaxy distribution, such as assembly bias \citep{HOD_Lin, Wechsler&Tinker, HOD_Xu, HOD_Barreira}, inspiring multiple extensions to the baseline HOD formalism. 

DESI employs a specific HOD model to generate its mock galaxy catalogs. In this work, we also explore variations in the HOD parameters to investigate how higher-order clustering statistics can improve constraints on the galaxy–halo connection. Following the methodology used in the construction of the official DESI mocks, we use the highly efficient \textsc{AbacusHOD} module to populate halo catalogs from the \textsc{AbacusSummit} simulations with mock galaxies. In addition to the baseline HOD prescription, the code supports a wide range of extensions, including galaxy assembly bias, velocity bias, and satellite profile flexibilities \citep{Yuan_2018, Abacus_HOD}. 
The \textsc{AbacusHOD} code is publicly distributed as part of the abacusutils package at \url{https://github.com/abacusorg/abacusutils}, with further documentation available at \url{https://abacusutils.readthedocs.io/en/latest/hod.html}.

\subsubsection{Baseline Model}
For LRG tracers, the baseline HOD model follows \cite{Zheng_2007}:
\begin{equation}
    \bar{n}_{\rm cent}(M)=\frac{f_{\rm ic}}{2}\Big(1+ \text{erf}\big[\frac{\log_{10}(M/M_{\rm cut})}{\sqrt{2}\sigma}\big]\Big),
\end{equation}
\begin{equation}
    \bar{n}_{\rm sat}(M)=\Big[\frac{M-\kappa M_{\rm cut}}{M_{1}}\Big]^{\alpha} \bar{n}_{\rm cent}(M),
\end{equation}
with five baseline parameters $M_{\rm cut}$, $M_{1}$, $\sigma$, $\alpha$, and $\kappa$.
$M_{\rm cut}$ is the characteristic minimum mass of halos that can host a central galaxy. $\sigma$ is the width of the cutoff profile and controls the steepness of the transition from $0$ to $1$ in the number of centrals. The number of satellite galaxies can be approximately described with a power law, where $M_{1}$ characterizes the typical halo mass that hosts one satellite galaxy and $\alpha$ is the power law index. $\kappa$, together with $M_{\rm cut}$, sets a mass threshold under which the probability of hosting a satellite is zero. 
We note that there is an extra term of $\bar{n}_{\rm cent}(M)$ in the satellite occupation function to ensure that a halo can only host satellites if there exists a central galaxy in it \citep{Guo_2015}. 

The DESI mocks have also included an incompleteness
parameter $f_{\rm ic}$, which controls the overall number density of the mock galaxies to match the observed galaxy number density. By definition, $0\leq f_{\rm ic}\leq 1$. 

\subsubsection{Extensions}
Beyond the baseline model, \textsc{AbacusHOD} supports multiple physically motivated HOD extensions, targeting diverse cosmological analyses. In this section, we briefly summarize several relevant extensions for the LRGs, which are employed in the construction of DESI mocks and may introduce variations in our analysis.

In addition to determining the expected number of galaxies per halo, the HOD framework also specifies the spatial and velocity distributions of galaxies within halos. Because DESI models the full 2D redshift-space correlation function $\xi(r_{p}, \pi)$, which is sensitive to galaxy velocities, a velocity bias model is incorporated. Following \cite{Guo_2015}, this model introduces two additional velocity bias parameters, $\alpha_{c}$ and $\alpha_{s}$, to allow the central and satellite velocities to deviate from the velocity of underlying dark matter, offering more flexibility in modeling redshift-space clustering.

With this extension, the HOD model for LRGs is fully specified with eight parameters: the five baseline parameters $M_{\rm cut}$, $M_{1}$, $\sigma$, $\alpha$, and $\kappa$; the incompleteness parameter $f_{\rm ic}$; and the velocity bias parameters $\alpha_{c}$ and $\alpha_{s}$. Redshift-space galaxy auto-correlation functions from the DESI One-Percent Survey (part of the Survey Validation from the Early Data Release) show that this model sufficiently describes LRG clustering, with no evidence for secondary biases or modifications from satellite profiles at the corresponding level of uncertainty \citep{one_percent}. As a result, DESI adopts this 8-parameter HOD model to construct the second-generation mocks, which we further describe in Section~\ref{sec-gen}. 

While the redshift-space 2PCF appears to be well described by the 8-parameter HOD model, we are interested in testing whether the more informative higher-order statistic, multiplet clustering, can further break degeneracies in HOD models that incorporate secondary biases. These biases introduce additional dependencies in the galaxy-halo connection beyond halo mass, potentially revealing more detailed aspects of galaxy formation physics.
\textsc{AbacusHOD} provides flexibility to include two forms of secondary bias: one based on halo concentration, and the other based on overdensity in the local environment. 
In the case of concentration-based bias, the HOD model incorporates the galaxy assembly history of the halo via the parameter of halo concentration \citep{assembly1, assembly2}. Alternatively, environment-based secondary bias introduces a dependence on the dark matter density around each halo, allowing galaxy occupation to vary with the local environment \citep{sec_bias}.

Following \cite{sec_bias}, \textsc{AbacusHOD}
models the biases by analytically mixing the secondary properties with the two mass parameters for centrals and satellites in the baseline model,
\begin{equation}
    \log_{10}M_{\rm cut}^{\rm mod} = \log_{10}M_{\rm cut}+A_{c}(c^{\rm rank}-0.5)+B_{c}(\delta^{\rm rank}-0.5),
\end{equation}
\begin{equation}
    \log_{10}M_{1}^{\rm mod} = \log_{10}M_{1}+A_{s}(c^{\rm rank}-0.5)+B_{s}(\delta^{\rm rank}-0.5),
\end{equation}
where $c$ denotes the halo concentration parameter and $\delta$ represents the overdensity in the local dark matter distribution. Both properties are ranked in narrow halo mass bins and normalized to $c^{\rm rank}$ and $\delta^{\rm rank}$, which range from $0$ to $1$. Then, the four parameters $(A_{c}, A_{s}, B_{c}, B_{s})$ quantify the strength of the secondary bias in the HOD model. Physically, a positive $A$ indicates a preference for lower-concentration halos, and a positive $B$ indicates a preference for halos in less dense environments. In the absence of secondary biases, all four parameters are zero. 

\subsection{DESI Second Generation Mocks}
\label{sec-gen}

The DESI second-generation mocks are constructed from the \textsc{AbacusSummit} N-body simulations, specifically the 25 realizations at the Planck 2018 fiducial $\Lambda$CDM cosmology \citep{Planck2018}. 
The simulation outputs are organized into discrete snapshots at different redshifts, and for each tracer, HOD models calibrated to the measured clustering from the One-Percent Survey are applied to simulate proper galaxy samples from these underlying dark matter density fields. 
For LRGs, tracers at $z<0.6$ are simulated using the periodic boxes at $z=0.5$, while those at $z > 0.6$ are modeled with the $z = 0.8$ snapshot. In both cases, HOD parameters are calibrated using the 8-parameter model described in Section~\ref{HOD_model}.
DESI then takes the best-fit HODs across different redshift snapshots and constructs ``cutsky'' mocks by converting coordinates in periodic boxes to celestial coordinates and redshifts. Redshift-space distortions are incorporated by adjusting the redshifts of simulated galaxies accordingly.

DESI has developed and applied various treatments to account for survey systematics in the mock catalogs. In this work, we focus on two types of mocks: one without fiber assignment and one with it implemented. The \texttt{complete} mocks replicate the footprint of Y3 observations and the target selection masks, but no additional fiber assignment choices are incorporated. Thus, these mocks are treated as complete samples, and all completeness weights are $1$. In contrast, the \texttt{altmtl} mocks are processed through the same fiber assignment pipeline applied for the actual observations, utilizing the real observational record and the same hardware status \citep{altmtl}. We expect these mocks to realistically reproduce survey systematics and enable unbiased clustering comparisons to observations.
Similar to the observational data, the mocks are organized into LSS catalogs, with corresponding random catalogs generated via the LSS pipeline. The resulting catalogs closely match the actual data in footprint, redshift distribution, and number density, enabling consistent application of any clustering analysis pipeline to both the mocks and the observations.

The second-generation mocks used in this work are the same as those employed in the DESI DR1 key projects \citep{DESI2024_v, DESI2024_iii}, but with the updated Y3 footprint. More details on the simulations and the construction of these mocks are available in \citet{2024mocks}, and we will further discuss our mock choices in Section~\ref{sec:calibration}.


%% file: calibration.tex

Because DESI is a fiber-based instrument, fiber assignment procedures and fiber collisions result in unavoidable incompleteness in our sample. For example, in DESI, if there is a high-redshift ($z>2.1$) quasar in the field, it will get priority during fiber assignment, and thus nearby LRGs will not be observed \citep{DESI1}. Also, as in other fiber-fed surveys, the physical size and number density of fibers limit the survey to record redshifts of galaxies too close to each other on the focal plane \citep{collision}. DESI fiber collision effects are even more complicated as the fibers are all robotically controlled. These effects have a non-trivial impact on clustering measurements. 
Especially as we are measuring galaxy-multiplet cross-correlations, this incompleteness in observations can cause missing objects in identified galaxy multiplets (i.e., a quad appears as a triplet, a triplet appears as a pair) and thus lead to bias in the higher-order statistics. 

Therefore, we present a direct empirical comparison of how the observations would appear without and with fiber assignment effects, using the \texttt{complete} and \texttt{altmtl} mocks. Specifically, we compute the correlation functions using each of the 25 \texttt{complete} and \texttt{altmtl} mocks and use the averages of the 25 realizations as benchmarks for modeled clustering without or with fiber assignment implementation.

\begin{figure}
    \centering
    \includegraphics[width=0.9\columnwidth]{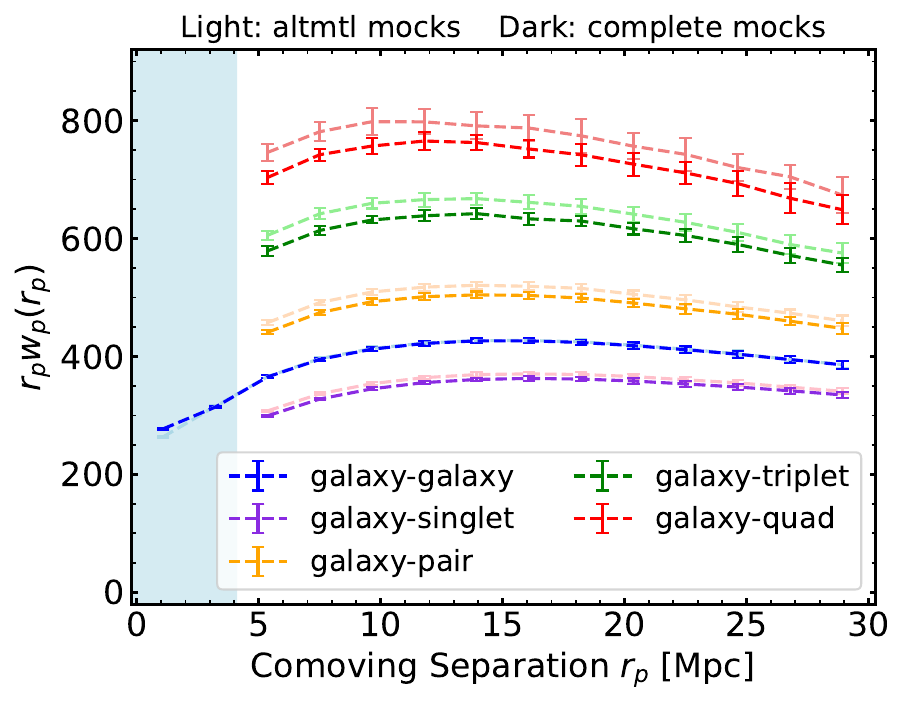}
    \caption{Comparison between the auto-correlations and the multiplet cross-correlations calculated from mocks without and with fiber assignment effects (the \texttt{complete} and \texttt{altmtl} mocks). The color differences follow the convention of this paper, while dark colors mark results from the \texttt{complete} mocks
    and light colors represent results from the \texttt{altmtl} mocks.}
    \label{fig:cross-corr-cali}
\end{figure}

Fig.~\ref{fig:cross-corr-cali} compares the auto-correlations and the cross-correlations with singlets, pairs, triplets, and quads calculated from the \texttt{complete} and \texttt{altmtl} mocks, with error bars representing the standard deviation across the 25 realizations. While the \texttt{altmtl} mocks exhibit a lower galaxy number density (approximately $88\%$) due to incomplete observations, the redshift distributions remain consistent between the \texttt{complete} and \texttt{altmtl} mocks, and the auto-correlation functions show good agreement on large scales. This consistency suggests that DESI's weighting scheme effectively calibrates survey systematics in auto-correlation measurements as expected.


Beyond the auto-correlations, the multiplet cross-correlations are expected to be more sensitive to survey geometry. Fig.~\ref{fig:cross-corr-cali} shows that the multiplet clustering amplitude inferred from the \texttt{complete} mocks is consistently lower than that measured with the \texttt{altmtl} mocks, indicating that when there is no additional assembly bias, a complete sample favors less massive dark matter halos. 
This makes sense because, in the \texttt{altmtl} mocks, fiber assignment can result in missing galaxies from a multiplet, causing the remaining system to exhibit larger clustering bias than multiplets of the same size in the \texttt{complete} mocks.
For example, a triplet from the \texttt{altmtl} mocks can potentially be a quad, which correlates preferentially with a more massive halo. Similarly, while a singlet in the \texttt{complete} mocks truly represents an isolated galaxy, a singlet in the \texttt{altmtl} mocks may arise from a higher-multiplicity system with unobserved members.

Because incomplete observations can artificially enhance the amplitudes of galaxy-multiplet cross-correlations, calibrating the impact of fiber assignment on multiplet clustering is necessary. As generating proper lightcones and cutsky mocks that match survey specifics is complicated and computationally expensive, this comparison between \texttt{complete} and \texttt{altmtl} mocks also serves as a nice empirical calibration for fiber assignment effects for later analysis in Section~\ref{sec:sec_bias}. We use the difference in clustering amplitudes, averaged over large scales ($r_p > 10\,\mathrm{Mpc}$), to correct for survey systematics before comparing observations with mocks generated from periodic boxes. While we understand this calibration is not perfect, as the second-generation mocks introduce their own systematics, it provides a simple benchmark to understand and isolate the effect of survey specifics, effectively eliminating contamination from survey geometry to the physics of galaxy-halo connections. 

%% file: hod.tex
To understand the physics of the galaxy-dark matter connection, we need to compare clustering statistics from observation and simulation. In this section, we present comparisons of multiplet clustering measured from the observations with those calculated from the DESI second-generation mocks and from mocks with modified halo occupation models. 

\subsection{Comparison between DESI mocks and observations}
\label{sec: comp}

The DESI second-generation mocks are constructed using the HOD model with best-fit parameters for the two-point correlation functions from the One-Percent Survey, while the \texttt{altmtl} mocks have been processed through the fiber assignment pipeline that closely replicates actual survey procedures. Therefore, these mocks offer robust theoretical predictions for galaxy clustering and are invaluable for interpreting DESI observations. Accordingly, we first compare the observed clustering with modeled predictions from the \texttt{altmtl} mocks, as they are supposed to incorporate both galaxy assembly physics and realistic survey systematics. 

We identify galaxy multiplets using galaxy catalogs from each of the 25 \texttt{altmtl} mocks and compute both the auto-correlation functions and the multiplet cross-correlation functions, following the methodology described in Section~\ref{sec:cluster}. As these mocks are organized into LSS catalogs with corresponding randoms, all pair counts are weighted in the same way as for galaxies from observations. For all clustering measurements, we use the average of the 25 \texttt{altmtl} mocks as the final prediction and compute the error bars as the standard deviation across the 25 realizations in each $r_p$ bin.

\begin{figure}
    \centering
    \includegraphics[width=0.9\columnwidth]{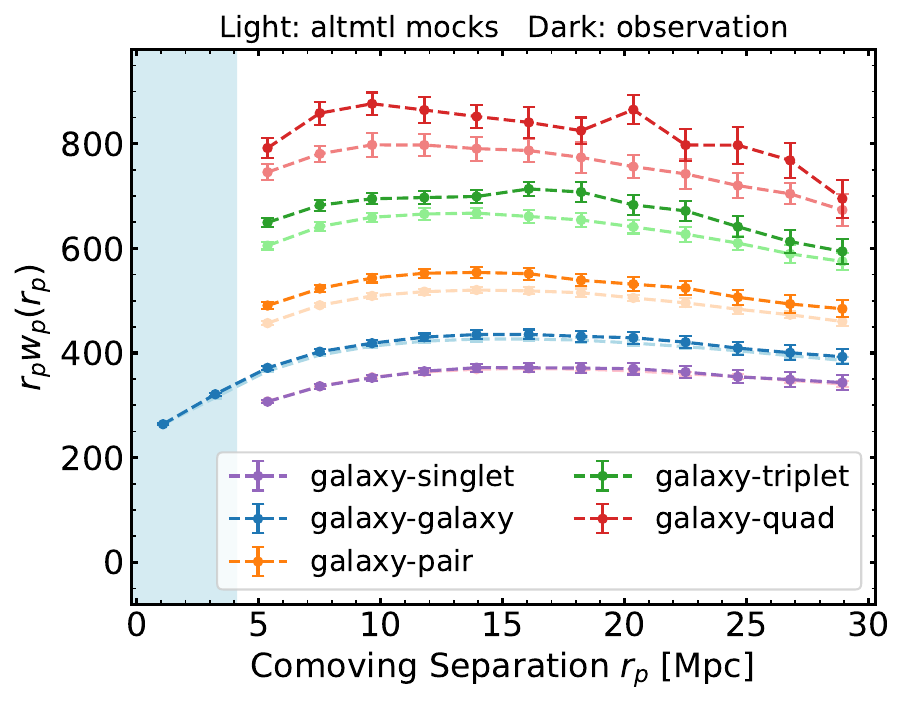}
    \caption{Comparison between the galaxy-multiplet cross-correlations measured from the Y3 observations and the \texttt{altmtl} mocks. Following Fig.~\ref{fig:cross-corr}, each color represents clustering measured with galaxy multiplets of different sizes, with lighter colors indicating the mock results and darker colors depicting the observations.}
    \label{fig:cross-corr-comp}
\end{figure}

Fig.~\ref{fig:cross-corr-comp} compares the multiplet cross-correlations calculated from observations and the \texttt{altmtl} mocks, including auto-correlations as a benchmark. Observational results are shown in darker solid lines, while those from the mocks are plotted in lighter colors.
We note that the error bars for the mocks differ from those of the observations: for the latter, we estimate uncertainties using jackknife resampling, while for the mocks, we use the scatter among independent realizations.
The plot shows that the auto-correlation from the mocks agrees reasonably well with the observations, indicating that the HOD model used in the second-generation mocks provides a good description of the projected correlation function and is reasonable to use in DESI cosmological analyses.
However, while the HOD model successfully reproduces a two-point correlation function relatively consistent with observations, this agreement does not extend to the cross-correlations with galaxy multiplets, where significant discrepancies appear. In particular, the galaxy multiplets in the mocks exhibit smaller clustering bias than those in the actual observations. If there is no additional assembly bias, this difference in the clustering amplitudes would suggest that mock multiplets tend to trace less massive dark matter halos; otherwise, there could be additional assembly bias specific to multiplets that alters clustering.


\begin{figure}
    \centering
    \includegraphics[width=0.9\columnwidth]{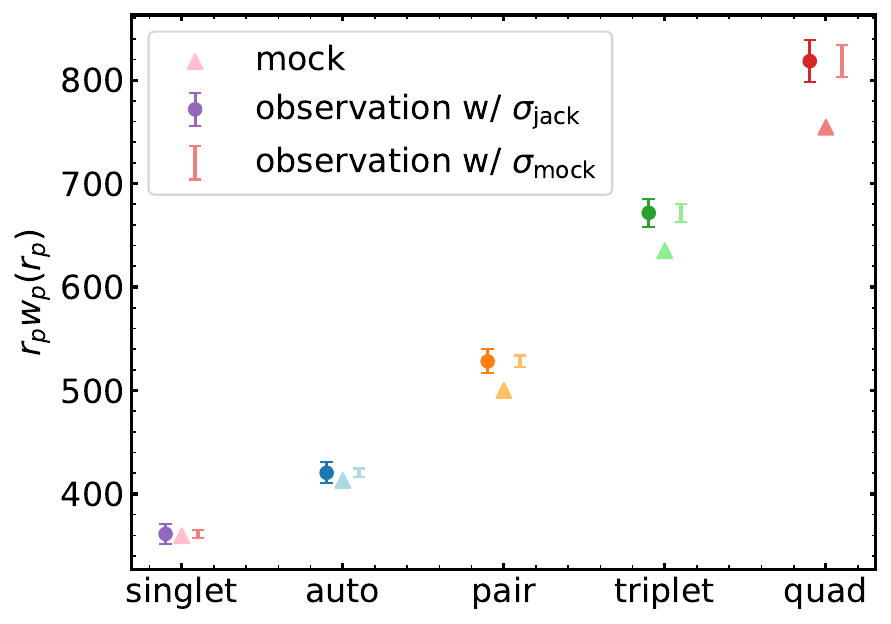}
    \caption{Comparison between the general clustering amplitudes for galaxies and multiplets, represented by the projected correlation functions averaged over a specific $r_{p}$ range. 
    Triangular points in lighter colors show results from the mocks, while darker circular points depict the observations. Two versions of error bars are included: jackknife errors estimated from the observations are shown on the left, and error bars capturing the scatter between mocks are on the right.}
    \label{fig:cross-corr-sum}
\end{figure}

To condense the information from the projected correlation functions and facilitate the comparison between observations and mocks, we average $r_{p} w_{p}$ over the scale $10\,\text{Mpc} < r_{p} < 30\,\text{Mpc}$ to obtain the general clustering amplitudes. Fig.~\ref{fig:cross-corr-sum} presents these condensed clustering statistics for both observations and mocks, including the auto-correlations as well as the cross-correlations with galaxy singlets, pairs, triplets, and quads. Here, we include two versions of error bars on the observations, one from jackknife resampling of the data and the other purely derived from the mocks. Specifically, $\sigma_{\rm jack}$ is calculated with the standard jackknife procedure (Equation 4) based on the average clustering amplitudes in the $50$ jackknife replicates described in Section~\ref{sec:cluster}, while $\sigma_{\rm mock}$ is obtained as the standard deviation of the 25 average clustering amplitudes measured from the 25 mocks.
We note that the jackknife error bars estimated with data consistently exceed the scatter between mocks, particularly for the cross-correlation with singlets and the galaxy auto-correlation. We repeat the same jackknife procedure on the mocks and find error bars of similar amplitudes to those from the real observations, suggesting that the larger variance is not due to unknown intrinsic fluctuations in the data but is instead a feature of the jackknife method itself. At the same time, the smaller scatter among the mocks suggests that the mock catalogs are not subject to additional sources of systematic uncertainty. Understanding that the jackknife method can be conservative, we include both the jackknife and the mock-based error bars to provide a more complete characterization of uncertainties in our measurements.

Despite the differences between the two error estimates, we observe statistically significant discrepancies at a few-sigma level between the mocks and observations for the multiplet cross-correlations, with multiplets in the simulations exhibiting smaller clustering amplitudes. The HOD fails to predict the higher-order correlations as effectively as it does for the auto-correlation. These discrepancies suggest that the vanilla HOD model is insufficient to describe the behavior of close galaxy multiplets, highlighting the potential of multiplet clustering as a higher-order statistic to offer better constraints on the galaxy-halo connection. 

We also directly compare the number densities of multiplets in the observations and mocks to complement the multiplet clustering analysis. The fraction of pairs among all galaxies in the mocks is about $5\%$ higher than in observations, the number density of triplets is about $1\%$ higher, and the observed number density of quads is consistent with statistical fluctuations in the mock predictions (about $1\%$). These discrepancies suggest that the mocks are slightly off-tuned, with the HOD performing less well for lower-mass halos. However, compared to the disagreement in clustering amplitudes, the differences in number densities are less significant and physically different, highlighting the distinct physics multiplet clustering probes and motivating additional assembly bias for the multiplets.

\subsection{Insights on Secondary Bias} \label{sec:sec_bias}
There have been studies showing that standard HOD models applied to cosmological N-body simulations may not be able to fully describe the galaxy distribution in hydrodynamical simulations that incorporate more complex physics of galaxy formation (i.e. \cite{boryana}).  
Nevertheless, employing N-body simulations and modeling the galaxy-halo connection with HOD is computationally efficient, and thus it is worth testing whether additional flexibility in the HOD can improve the performance of this simple but powerful approach. At the same time, the various discrepancies between observations and mocks in different multiplet clustering measurements are reassuring that these higher-order cross-correlations can provide more information about how galaxies trace dark matter. 

As an efficient higher-order statistic, multiplet clustering offers valuable opportunities to study the effect of secondary halo properties on the galaxy-halo connection. 
While in this paper, we will not attempt to constrain the HOD parameters through full statistical inferences, this section demonstrates how this novel statistic can break degeneracies in galaxy-halo connection models and provide insights into secondary biases through testing the reconstruction of multiplet cross-correlations under various modifications to the vanilla HOD model.

We utilize the cubic box at Planck 2018 cosmology from the \textsc{AbacusSummit} simulation suite, specifically the redshift snapshot at $z=0.5$, as the model template for the underlying dark matter density field. We then generate mock galaxy catalogs with different HOD prescriptions using the \textsc{AbacusHOD} package and incorporate redshift-space distortions along the $z$-axis.
The incompleteness parameter, $f_{\rm ic}$, is tuned to match the number density with that of the \texttt{complete} mocks.
To consistently identify galaxy pairs and multiplets in the periodic box as in real observations, we use the $z$-axis as the LOS and apply periodic wrapping conditions to ensure that galaxies interact with each other in a continuous system. We set the separation threshold to $10\,\rm Mpc$ in 3D and $1.5\,\rm Mpc$ in the $x$-$y$ plane for the selection of galaxy pairs, which are linked into multiplets using the friends-of-friends algorithm. 
These multiplets are then used to compute the cross-correlation functions, and we calibrate the average clustering amplitudes following Section~\ref{sec:calibration} before comparing with real observations. 

Although the projected correlation function $w_p$ is largely insensitive to velocity bias or satellite profile parameters, these factors could potentially affect multiplet identification due to redshift space distortions. Therefore, before exploring assembly bias, we test whether satellite velocity and spatial biases can alter multiplet clustering signals and improve the agreement with observations.

The second-generation mocks are generated with the inclusion of satellite velocity bias, carefully calibrated to match the full-shape correlation function $\xi(r_p, \pi)$ down to small scales. 
Thus, the fact that the observed clustering of multiplets differs from the mocks already indicates that the velocity bias alone might be insufficient to explain the discrepancy identified in Section~\ref{sec: comp}. 
To further support this conclusion and explore the potential impact of satellite velocity bias, we fix the five baseline HOD parameters and vary the satellite velocity bias parameter $\alpha_s$ across a wide range from $0.5$ to $1.2$. While this range spans physically very distinct regimes, well beyond the typical values measured from DESI, similar samples, and hydrodynamical simulations \citep{SDSS_v_bias, TNG_v_bias}, it induces only mild variations in multiplet clustering amplitudes, with all differences remaining below the $1\sigma$ level. Moreover, we find that the number density of multiplets is more sensitive to $\alpha_s$ than the clustering amplitudes are, suggesting that satellite velocity bias can be well constrained through careful calibrations of number densities in addition to small-scale full-shape clustering.
Thus, the influence of satellite velocity bias is effectively limited for our analysis.

The case for satellite spatial bias is more straightforward. Varying the satellite spatial bias parameter $s$ from $-0.3$ to $0.3$ results in only minor variations in the amplitudes of multiplet clustering, allowing us to safely exclude satellite spatial bias from further consideration.

We remind that these tests do not necessarily imply that multiplet clustering lacks constraining power for velocity bias or satellite profiles. In fact, we expect that 2D full-shape cross-correlation functions with multiplets can encode valuable information about these parameters. However, such analyses are beyond the scope of this work, as we are presenting a highly compressed statistic, the average clustering amplitude from the projected correlation function, to provide simple demonstrations of potential future applications. Since assembly bias is important for the galaxy-halo connection and early DESI data have limited statistical power to distinguish between models with different assembly biases, we choose to demonstrate the impact of multiplet clustering specifically on the secondary biases.

We focus on two physically motivated HOD extensions that capture the impact of halo assembly history and local environment on the galaxy-halo connection, following the implementation described in Section~\ref{HOD_model}. To ensure consistency, we first tune a vanilla HOD model to match the two-point correlation function of the \texttt{complete} mocks, and refer to this model as the ``baseline'' throughout the remainder of this section. 

\begin{figure}
    \centering
    \includegraphics[width=0.95\columnwidth]{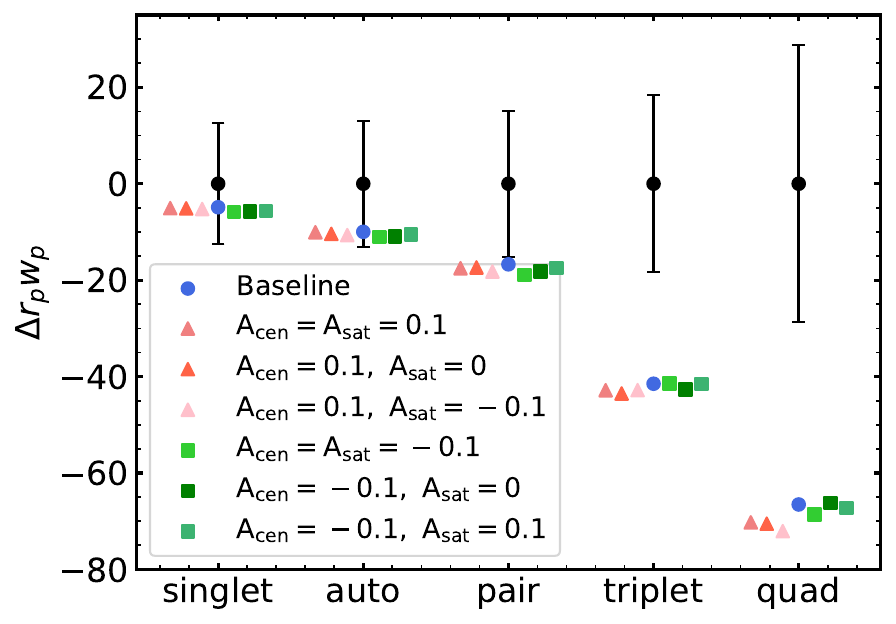}
    \caption{The average clustering amplitudes measured from observations and predicted by HOD models with different concentration-based secondary biases added to the same set of baseline parameters.
    Observational data points with jackknife error bars are fixed at zero as references. Models with different secondary biases, as indicated in the labels, are plotted from left to right on either side of the baseline model results.}
    \label{fig:A_test}
\end{figure}

First, we experiment with the concentration-based secondary bias described by $A_{c}$ and $A_{s}$. Fig.~\ref{fig:A_test} showcases the deviation in the two-point auto-correlations and the multiplet cross-correlations with the inclusion of different $(A_{c},A_s)$ on top of the baseline parameters.
We compare the differences in clustering amplitudes ($\Delta r_p w_p$) between the models, including the baseline and those with concentration-based biases, and the observations, which are centered at zero with jackknife error bars indicating uncertainty.
The plot demonstrates that the inclusion of $(A_{c}, A_s)$ on top of the same set of baseline HOD parameters does not significantly affect the auto-correlation, suggesting that the two-point function alone is not sensitive enough to distinguish between models with different concentration-based biases. 
The impact of $(A_{c},A_s)$ on higher-order cross-correlations is larger. However, compared with the increasing error bars for correlations with larger multiplets and the discrepancy between mocks and observations, the improvement in the differentiating power of multiplet clustering does not seem statistically important.  

\begin{figure}
    \centering
    \includegraphics[width=0.95\columnwidth]{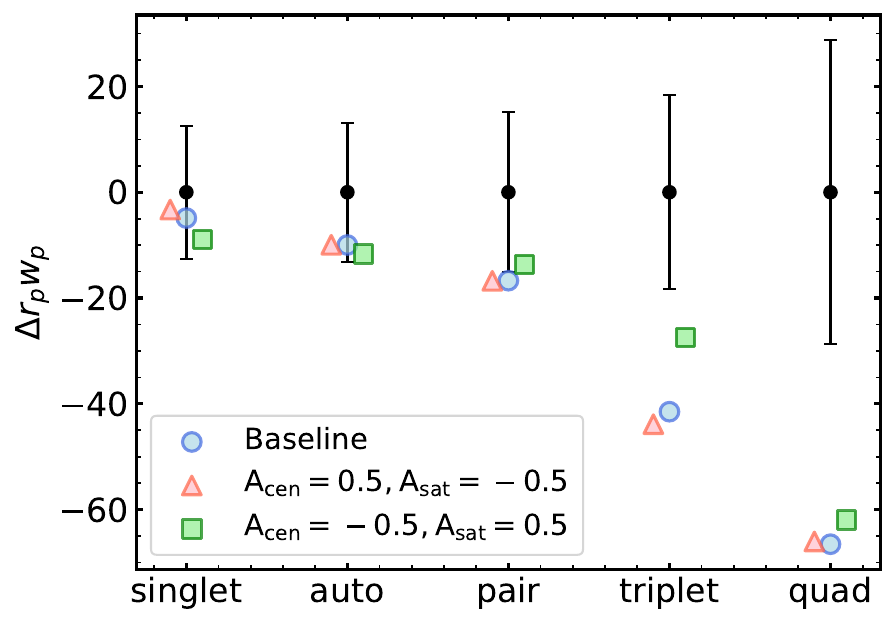}
    \caption{Example of average clustering amplitudes measured from observations and predicted by HOD models incorporating different concentration-based secondary biases of enhanced magnitudes. Observational data points with jackknife error bars are fixed at zero as references. Predictions from the baseline HOD model and from models incorporating positive and negative $A_{c}$ are represented by circles, triangles, and squares, respectively.}
    \label{fig:A_test2}
\end{figure}

We further increase the magnitudes of the secondary bias parameters $A_{c}$ and $A_s$ to induce a larger deviation from the auto-correlation predicted by the baseline model, thereby enhancing the dependence on halo concentration and ensuring their statistical significance. 
We then perform a systematic grid search over the five vanilla HOD parameters around their baseline values at two sets of fixed $(A_c, A_s)$, selecting the combination of parameters that best fits the baseline two-point function based on the $\chi^2$ calculated using the jackknife error bars from observations.
The resulting shifts in vanilla HOD parameters remain small after introducing a more prominent dependence on halo concentration, and Fig.~\ref{fig:A_test2} illustrates the modifications of multiplet clustering with the inclusion of these stronger concentration-based biases. While the agreement between the model and observations can be improved, the overall conclusion remains that the impact of concentration-based secondary bias is not statistically significant enough and including $(A_{c}, A_s)$ does not resolve the discrepancies observed in higher-order clustering.

Another set of secondary biases we explore is the environment-based bias described by $B_{c}$ and $B_{s}$. Again, we start by including different values of $(B_c,B_s)$ on top of the same set of baseline HOD parameters. However, this time we observe large deviations in the predictions of auto-correlations from different models, which is consistent with previous work that vanilla HOD parameters are more sensitive to the inclusion of environment-based bias. To recover a reasonable fit to the two-point function so that the comparison of multiplet clustering predictions is meaningful, we again perform a systematic grid search around the baseline HOD parameters at fixed values of $(B_c, B_s)$. We explore two simplified scenarios, $B_c=B_s=\pm 0.05$, and terminate the grid search once the minimum $\chi^2$ calculated with jackknife error bars falls below a certain threshold. Following this procedure, we use the best parameters from the grid search to generate two HOD models with different environment-based secondary biases, while both models predict reasonable galaxy-galaxy auto-correlations. Moreover, as we modify the bias parameters systematically, we can infer the relation between HOD parameters.  
For example, when including negative $(B_{c}, B_{s})$, we see an increase in $\sigma$ and a decrease in $M_{1}$ compared to the original baseline parameters, which translates to moving galaxies to lower mass halos in denser environments and is consistent with previous work based on complete statistical inference \citep{Abacus_HOD}. 

\begin{figure}
    \centering
    \includegraphics[width=0.95\columnwidth]{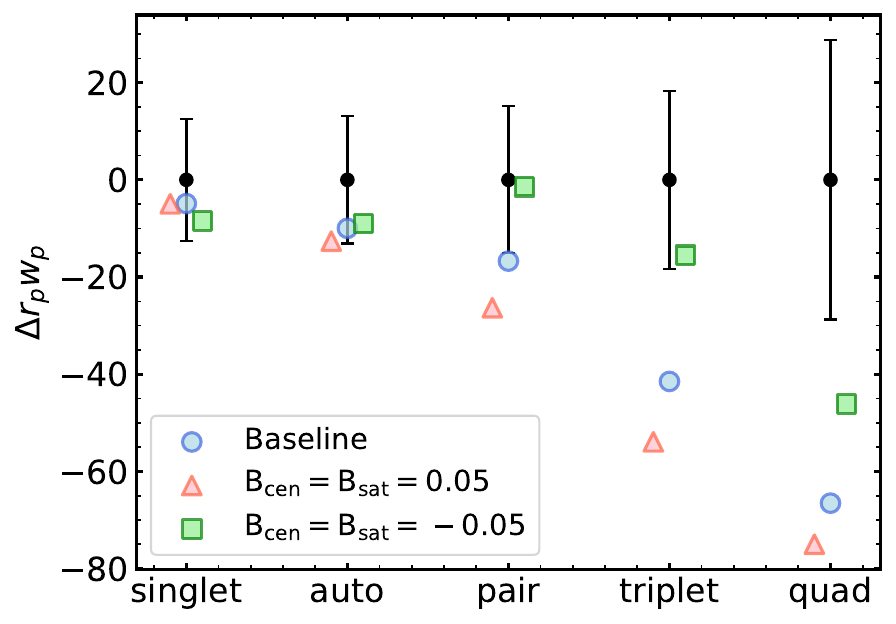}
    \caption{Similar to Fig.~\ref{fig:A_test2} but for different environment-based secondary biases $(B_{c}, B_{s})$.}
    \label{fig:B_test}
\end{figure}

Given that these models can fit the observed autocorrelation equally well, we treat them as benchmarks and evaluate their performance using higher-order statistics. Fig.~\ref{fig:B_test} compares the averaged auto- and cross-correlations from observations and mocks incorporating different environment-based secondary biases. While all three models provide reasonable fits to the auto-correlation, their performance diverges when describing the galaxy-multiplet cross-correlations. We find that these deviations in multiplet clustering predictions from different environment-based biases are much more prominent than those from different concentration-based biases. Moreover, including positive or negative environment-based biases shifts the amplitudes of the cross-correlations in opposite directions. Notably, the model with negative $(B_c, B_s)$ significantly improves the consistency between mocks and observations for the multiplet cross-correlations. Therefore, we suggest that environment-based secondary bias is potentially important for fully capturing the physics of the galaxy-halo connection and that multiplet clustering offers a unique opportunity to reveal this preference.

It is worth noting that the results presented in this section are primarily methodology-oriented. Our aim is not to constrain the HOD parameters directly, but rather to provide insights into the use of higher-order statistics and the inclusion of secondary biases when modeling galaxy clustering.
We demonstrate that incorporating galaxy assembly bias into the HOD model, particularly environment-based secondary biases, has the potential to provide a more accurate description of the measured galaxy-multiplet clustering. Although definitive physical conclusions would require a full statistical inference on the data, we observe a trend that LRGs favorably occupy halos in denser environments, as indicated by the preference for negative environment-based secondary bias.
Furthermore, we highlight the potential of utilizing multiplet clustering to enhance the constraints on galaxy-dark matter connections. When models produce similar predictions for the two-point auto-correlation, making them difficult to distinguish due to statistical uncertainties, incorporating galaxy-multiplet cross-correlations in the analysis is helpful to break model degeneracies. As DESI continues to survey larger sky areas with higher completeness, the errors in this analysis are expected to decrease significantly, further strengthening the constraining power of multiplet clustering.

%% file: dis.tex
In this work, we propose multiplet clustering as an efficient higher-order statistic for analyzing galaxy clustering. Using LRGs from DESI DR2, we identify small groups of nearby galaxies, referred to as multiplets, and cross-correlate the resulting multiplet field with the general galaxy field to measure galaxy-multiplet clustering. While this estimator is computationally much cheaper than conventional higher-order statistics such as the 3PCF or bispectrum, it still captures higher-order information and small-scale nonlinearities, offering sensitivity to galaxy bias and cosmology.
Specifically, we cross-correlate galaxy singlets, pairs, triplets, and quads with the general galaxy field to obtain the galaxy-multiplet cross-correlation functions. We observe larger clustering bias for larger multiplets as tracers of more massive dark matter halos, suggesting that these multiplets trace the large-scale structure differently than individual galaxies. 

We compare the observed clustering statistics with mock catalogs constructed using the cosmological N-body simulation \textsc{AbacusSummit} with HOD models describing the galaxy-halo connection. While the auto-correlation measured with the official DESI second-generation mocks agrees well with actual observations, the multiplet cross-correlations from the data exhibit higher amplitudes compared with those predicted by the mocks (highlighted in Fig.~\ref{fig:cross-corr-comp} and Fig.~\ref{fig:cross-corr-sum}). 
This discrepancy indicates that, following the simplest assumptions, the multiplets in the mocks preferentially occupy less massive halos than those from observations, or that there could be additional assembly bias specific to the multiplets, suggesting the importance of higher-order statistics in capturing the full complexity of the galaxy-halo connection and motivating the exploration of extended HOD models that include secondary biases.
We test two physically motivated HOD extensions dependent on secondary halo properties: halo concentration and local environment. While these models are calibrated to match the same two-point correlation function, they produce different predictions for the galaxy-multiplet cross-correlations. We showcase that including secondary biases can potentially describe the higher-order correlations better, and the HOD model with environment-based secondary bias generates the most significant improvement in the data-model agreement, indicating that galaxies prefer to occupy halos in denser environments. 

We highlight the potential of using multiplet clustering to break degeneracies in models describing galaxy bias. Even though we have not performed a full statistical inference on the data, we observe that different models with the same descriptive power for the auto-correlation diverge in their predictions for the galaxy-multiplet cross-correlations. Therefore, the employment of these higher-order correlation functions, beyond providing information on how galaxy multiplets of various sizes trace dark matter differently, offers valuable opportunities to distinguish between complex models of the galaxy-dark matter connection. 

While we establish a good foundation for multiplet clustering analysis, the physical conclusions for the galaxy-halo connection are not yet comprehensive because the HOD modeling presented in this work does not utilize a quantitative likelihood analysis, which is essential for comparing HOD models and constraining the parameters statistically. Based on simple observations of various performances of HODs with different model extensions, we anticipate that a complete statistical inference of HOD models utilizing the multiplet clustering statistics, along with careful covariance matrix calculations and a larger data sample from future DESI observations, will provide powerful insights into galaxy assembly bias and potentially cosmology. Furthermore, we plan to investigate the full-shape correlation as it offers more information in the redshift space. We also hope to extend the analysis to other classes of galaxies beyond LRGs, considering their distinct physical properties. 

Beyond clustering analysis, the multiplets themselves encode interesting physics and offer opportunities for many other cosmological studies. For example, we can investigate how halo mass correlates with multiplet properties, such as internal separations in real space and redshift space, to further understand how galaxy multiplets of various properties trace the large-scale structure differently. Moreover, because isolated galaxies and galaxies residing in multiplets are tracers of the same underlying matter distribution but have different biases, they are potentially useful for multi-tracer cosmology analysis \citep{multi-tracer}, where sample variance can be effectively suppressed, allowing for more precise measurements of redshift-space distortions. 

We can also stack weak lensing signals around the multiplets and compare them to galaxy-galaxy lensing measurements to probe the matter distribution of the universe. While joint analyses between galaxy clustering and galaxy-galaxy lensing have proven valuable for reconstructing galaxy bias \citep{lensing-hod1, lensing-hod2, lensing-cos}, extending such analyses to another class of objects, such as galaxy multiplets, can further enhance our understanding of the galaxy-dark matter connection and improve constraints on cosmological parameters such as the amplitude of matter clustering $\sigma_{8}$. Specifically, adding lensing information allows us to calibrate the matter distribution around multiplets and verify the source of their enhanced clustering signals.
Moreover, an observational tension exists between galaxy clustering and galaxy-galaxy lensing \citep{tension}, and \cite{solve-tension} argues that incorporating galaxy assembly bias into the HOD framework can help explain this lensing tension. With extra information provided by the multiplet statistics, we can further evaluate this tension and explore possible solutions. 

Finally, we would like to explore cross-correlations with surveys of the Sunyaev-Zeldovich effect as another probe of the density distribution of the universe, while investigating the thermal and dynamical properties of the multiplets. 
